\documentclass[preprint2]{aastex}

\usepackage{epsfig}

\def\ga{\mathrel{\m

athchoice {\vcenter{\offinterlineskip\halign{\hfil%
$\displaystyle##$\hfil\cr>\cr\sim\cr}}}{\vcenter{\offinterlineskip%
\halign{\hfil$\textstyle##$\hfil\cr>\cr\sim\cr}}}{\vcenter{%
\offinterlineskip\halign{\hfil$\scriptstyle##$\hfil\cr>\cr\sim\cr}}}%
{\vcenter{\offinterlineskip\halign{\hfil$\scriptscriptstyle##$\hfil\cr%
>\cr\sim\cr}}}}}
\let\gsim=\ga

\def\la{\mathrel{\mathchoice {\vcenter{\offinterlineskip\halign{\hfil%
$\displaystyle##$\hfil\cr<\cr\sim\cr}}}{\vcenter{\offinterlineskip%
\halign{\hfil$\textstyle##$\hfil\cr<\cr\sim\cr}}}{\vcenter{%
\offinterlineskip\halign{\hfil$\scriptstyle##$\hfil\cr<\cr\sim\cr}}}%
{\vcenter{\offinterlineskip\halign{\hfil$\scriptscriptstyle##$\hfil\cr%
<\cr\sim\cr}}}}}
\let\lsim=\la

\def\gsim{\mathrel{\rlap{\lower 4pt \hbox{\hskip 1pt $\sim$}}\raise 1pt
\hbox {$>$}}}
\def\lsim{\mathrel{\rlap{\lower 4pt \hbox{\hskip 1pt $\sim$}}\raise 1pt
\hbox {$<$}}}

\begin{document}


\title{Evolution of Very Massive Population III Stars with Mass Accretion from Pre-Main Sequence to Collapse}

\author{Takuya Ohkubo\altaffilmark{1,2}, Ken'ichi Nomoto\altaffilmark{2,1}, Hideyuki Umeda\altaffilmark{1}, \\
Naoki Yoshida\altaffilmark{2,3}, and Sachiko Tsuruta\altaffilmark{4}} 
 

\altaffiltext{1}{Department of Astronomy, School of Science, University of Tokyo, 7-3-1 Hongo, 
Bunkyo-ku, Tokyo 113-0033, Japan; ohkubo@astron.s.u-tokyo.ac.jp, umeda@astron.s.u-tokyo.ac.jp} 
\altaffiltext{2}{Institute for the Physics and Mathematics of the Universe (IPMU), University of Tokyo, 5-1-5 Kashiwanoha, 
Kashiwa, Chiba 277-8568, Japan; nomoto@astron.s.u-tokyo.ac.jp, naoki.yoshida@ipmu.jp} 
\altaffiltext{3}{Department of Physics, Nagoya University, Furocho, 
Chikusa-ku, Nagoya 464-8602, Japan} 
\altaffiltext{4}{Department of Physics, Montana State University, Bozeman, MT 59717-3840; 
uphst@gemini.msu.montana.edu} 

\begin{abstract}
We calculate the evolution of zero-metallicity Population III (Pop III) stars 
whose mass grows from the initial mass of $\sim 1M_{\odot}$ by accreting the 
surrounding gases. Our calculations cover a whole evolutionary stages from 
the pre-main sequence, via various nuclear burning
stages, through the final core collapse or pair-creation instability phases. 
We adopt the following stellar mass-dependent accretion rates which 
are derived from cosmological simulations of early structure formation
based on the low mass dark matter halos at redshifts $z \sim 20$: 
(1) the accretion rates for the first generation (Pop III.1) stars and 
(2) the rates for zero-metallicity but the second generation 
(Pop III.2) stars which are affected by radiation from the Pop III.1 
stars. For comparison, we also study the evolution with 
the mass-dependent accretion 
rates which are affected by radiatibe feedback.
We show that the final mass 
of Pop III.1 stars can be as large as $\sim 1000M_{\odot}$, 
beyond the mass range ($140 - 300M_{\odot}$) 
for the pair-instability supernovae. 
Such massive stars undergo core-collapse to form intermediate-mass 
black holes, 
which may be the seeds for merger trees to supermassive black holes. 
On the other hand, Pop III.2 stars become less massive 
($\lsim 40 - 60M_{\odot}$), being in the mass range of ordinary 
iron core-collapse stars. 
Such stars explode and eject heavy elements to contribute to chemical 
enrichment of the early universe as observed in the abundance patterns  
of extremely metal-poor stars in the Galactic halo. 
\end{abstract}

\keywords{accretion, accretion disks -- nuclear reactions, nucleosynthesis, abundances -- stars: abundances -- stars: evolution -- stars: formation -- supernovae: general}

\section{INTRODUCTION}

Just after the Big Bang,
a cosmic primordial gas consists mostly of H, He and a small amount of light
elements (Li, Be, B, etc). The first heavier elements, such as C, O, Ne, Mg,
Si and Fe, must be synthesized during the evolution of the first
(metal-free $=$ Population
III $=$ Pop III) stars early in the history of the universe.
It is generally thought that massive
Pop III stars distribute synthesized heavy elements by supernova 
explosions (e.g., \citealt{Nomo06}). 

If Pop III stars are sufficiantly massive, they could also form
intermediate mass black holes (IMBHs) ($\sim 5 \times
10^{2-5}M_{\odot}$). The formation of IMBHs would have important
implications. Stellar mass black holes ($\sim 10M_{\odot}$) are
formed as the central compact remnants of ordinary massive (25 -
140$M_{\odot}$) stars at the end of their evolution. Supermassive
black holes (SMBHs) ($\sim {10}^5$ - ${10}^9 M_{\odot}$) are now
known to exist in the center of almost all galaxies (e.g.,
\citealt{Korm95, Bend05}), but their formation processes are
largely unknown. Recently, the role of IMBHs caught much attention
in the context of the formation of SMBHs. One of the viable
scenarios for SMBH formation is a merger tree model in which seed
black holes of a few $\times 10^2M_{\odot}$ formed early (e.g., at
$z \sim 20$) are assumed to have merged and grown to become SMBHs
(e.g., \citealt{Mada01, Volo03}). In this model, the small seed
blackholes generally go through IMBH stages at some intermediate
epochs, e.g., between $z \sim 20$ and 0.

For both chemical enrichment and the formation of IMBHs in the
early universe, the massive Pop III stars may play important
roles. In order to clarify the fate of Pop III stars which depends
sensitively on the stellar mass, $M$, we study the evolution of
mass accreting Pop III stars. In the present paper, we generally
call stars with mass $M > 100M_\odot$ very massive stars (VMSs).
Among VMSs, we refer to stars with $M > 300M_\odot$ as
core-collapse very massive stars (CVMSs) (\citealt{Ohku06}).

The standard cosmological model based on dark energy and cold dark
matter (CDM), the so-called $\Lambda$CDM model, posits that
structure forms hierarchically from smaller mass objects (a
bottom-up scenario; \citealt{Davi85, Ostr03, Kirs03}). Small density 
fluctuations in the early universe seed nonlinear growth of
structure via gravitational instability. According to this model,
smaller, stellar size objects formed first and then larger
structures such as galaxies are assembled by merging of these
smaller units. The first generation stars are predicted to be
formed when the age of the universe was less than a few hunred
million years (\citealt{Couc86, Tegm97, Yosh03}), whereas,
observationally, the WMAP data suggests that it is $z \sim 10$
(\citealt{Sanc06, Sper07}).

In the early universe, dark matter plays an important role for the formation
of the first generation stars. Dark matter (DM) gathers due to gravity and
forms minihalos with typical mass of $\sim 10^6M_\odot$
(\citealt{Haim96, Tegm97, Full00}). After the formation of minihalos,
the baryon gas gravitationally collapses to form a star. Since the
primodial gas contains no heavy elements, molecular hydrogen
is the only efficient coolant indispensable for star formation
(\citealt{Peeb68, Mats69, Pall83}).

Theoretical studies on star formation suggest that the initial
mass function (IMF) of Pop III first stars may be different from
the present one, possibly dominated by very massive stars (e.g.,
\citealt{Omuk98, Naka99, Abel02, Brom99, Omuk03}). A
number of numerical simulations of the formation of the first
stars have been carried out. They are, either one-dimension
calculations (\citealt{Haim96, Naka02, Ahn07}) or three
dimensional simulations (\citealt{Abel00, Abel02, Brom99, 
Brom03a, Full00, Yosh06, OShe06a, OShe06b, Gao07}). These studies
generally suggest that the typical mass of a primodial gas cloud
is $\sim 10^3 - 10^5M_\odot$. The latest detailed calculations
have shown that a protostellar core of as small as $M \sim
10^{-2}M_\odot$ is formed at the center of a primordial gas cloud
(\citealt{Yosh08}). After such a small proto-stellar core is
formed, the gas surrounding it accretes on the protostar and the
stellar mass increases through the pre-main sequence stage.



A crucial parameter is the mass accretion rate, which determines the typical
mass of Pop III stars (e.g. \citealt{Tan04}). Cosmological
simulations (e.g. Abel et al.2002; \citealt{Brom04, Yosh06, Gao07})
have shown that the accretion rate is as high as ${\dot M} \sim
10^{-2}M_\odot \; \rm{yr^{-1}}$ when the core mass is small ($M \lsim
10M_\odot$) and it decreases with increasing mass. These 
studies have suggested that first stars might be more massive than
$100M_\odot$, in the zero-metallicity environment.
\cite{Omuk03} calculated protostar evolution with
constant accretion rates. They treated the accretion rate
as a free parameter and set it to be of the order of
${\dot M} \sim 10^{-2}M_\odot \; {\rm yr^{-1}}$ and obtained that
the final mass could exceed
$100M_\odot$. However, they also found that, if ${\dot M} >
4 \times 10^{-3}M_\odot \; \rm{yr^{-1}}$, the stellar radius rapidly 
expands to prevent further growth. Thus the final fate could not be 
predicted. If mass accretion continues through its lifetime ($\sim$ a few 
$\times 10^6$ years), i.e., if mass accretion is not impeded by 
feedback from the star itself, the final mass may reach several
times $10^2M_{\odot}$ or more.

Stars formed in the way described above
(which are called `Pop III.1' stars; \citealt{McKe08}) radiate a large amount of UV photons
and build up HII regions with a few kiloparsec diameter
(\citealt{Yosh07}). This environment promotes the formation of HD
molecules, which can cool the gas down to as low as $40 - 50$K.
In a gas cloud at such low temperatures,
typical stellar mass formed is $\sim 40M_{\odot}$, smaller than Pop III.1
stars. Stars formed in such environment are called `Pop III.2' stars
(\citealt{McKe08, John08}).

The possibility of the formation of VMSs as Pop III.1 stars has renewed
the interest in the final fate of such VMSs. Previous studies,
which did not take into account mass accretion,
have shown that the final fate of the VMSs is sensitive to the
stellar mass $M$ (e.g., \citealt{Bark67,
Raka68, Ober83, Bond84, Arne96}). If $M > 300M_{\odot}$, the stars undergo
core-collapse to form IMBHs, which we call CVMS.
If $M \simeq 140 - 300M_{\odot}$, the stars
undergo pair instability supernovae (PISNe) and disrupt completely,
ejecting a large amount of heavy elements.
Thus PISNe have been suggested to be the
main source of chemical enrichment in the early universe. However,
recent detailed comparisons between the observations of extremely
metal poor (EMP) stars \citep{Cayr04} and the nucleosynthesis yields of
PISN models \citep{Umed02, Hege02} have shown that the PISN yields are
hard to reproduce the abundance patterns of EMP stars \citep{Cayr04}.

It is interesting to examine theoretically under what conditions the
Pop III stars end their lives as PISNe.
The related question of whether CVMSs ($M \sim 300 -
10^5M_{\odot}$) could actually form is of great importance, for
instance, to understand the origin of IMBHs.
For this purpose, in this paper
we adopt the realistic mass accretion rate obtained by \cite{Yosh06},
which follows the evolution
of dense gas clumps
formed at the centers of rather low mass ($\sim
10^{5-6}M_{\odot}$) dark matter halos at redshifts $z \simeq 20$.



In the present paper, we calculate stellar evolution of Pop III
stars growing by accretion and investigate the final fate of such
stars. So far all previous calculations of protostar evolution
ended at the onset of hydrogen burning, at the start of the
main sequence, and hence they do not answer the question of how
such massive Pop III objects evolve after they start hydrogen
burning and what the final stellar mass can be. Therefore, the
major purpose of the current studies is not only to follow the
pre-main sequence evolution studied by \cite{Omuk03} and
\cite{Yosh06}, but also to continue further stellar
evolution by adopting the mass accretion rate obtained by the
cosmological simulation \citep{Yosh06}. We calculate whole nuclear
burning stages, until the point where the star ends its life
with core collapse and/or explosion.

Following the introduction in this section, our models are
described in Section 2 and our results are presented in Section 3.
In the last two sections, 4 and 5, we give discussion, summary and
concluding remarks.

\section{MODELS IN PRESENT STUDIES}

In Ohkubo et al. (2006), we
calculated evolution, nucleosynthesis, explosion and collapse of
Pop III CVMSs starting from the main-sequence in the absence
of mass accretion. More
realistically Pop III stars are formed along with cosmological
structure formation, and the mass increases by accretion and the
final mass may be in the range of VMS ($M > 100M_{\odot}$).
In the present study, we calculate the
evolution of Pop III stars with mass accretion, starting from the
pre main-sequence phase with small mass comparable to the solar mass,
until core collapse or pair instabillity explosion. In this
subsection we summarize our models and assumptions at the initial
and each subsequent stage.

To calculate presupernova evolution through the early 
hydrodynamical phase, we adopt the Henyey-type hydrodynamical 
stellar evolution  
code \citep{Nomo82, Nomo88, Umed99, Umed02, Umed05}. 
The mass accretion is calculated 
with the method of \cite{Neo76} and \cite{Nomo82}. We adopt the nuclear 
reaction network developed by \cite{Hix96} for calculating 
nucleosynthesis and
energy generation at each stage of the evolution. 
We include 51 isotopes up to Si until He burning 
ends, and 240 up to Ge afterwards. 

We start with a $1.5M_{\odot}$ pre main-sequence star, a typical
mass size in the low-mass star ranges. We investigate how this low
mass stellar core grows up to a massive one with gas accretion.
The starting mass of $1.5M_{\odot}$ is larger than the initial
mass set by \cite{Omuk03}, $0.01M_{\odot}$, by more than two
orders of magnitude. However, the time it takes for the star to
increase its mass from $0.01M_{\odot}$ to $1M_{\odot}$ is $\sim
10^2$ yr. This is negligible compared with the overall lifetime of
a star. Physically, convection over the whole star occurs in the
contracting phase. So the starting point little affects the later
evolution, and as one can see later, we can qualitatively
reproduce the protostellar evolution shown in \cite{Omuk03}. Since
we consider Pop III stars the chemical composition chosen are
$X({\rm H}) = 0.753, X({\rm D}) = 2 \times 10^{-5}, X(^3{\rm He})
= 2 \times 10^{-5}, X(^4{\rm He}) = 0.247$, and $X(^7{\rm Li}) = 2
\times 10^{-10}$.

\begin{figure}
\plotone{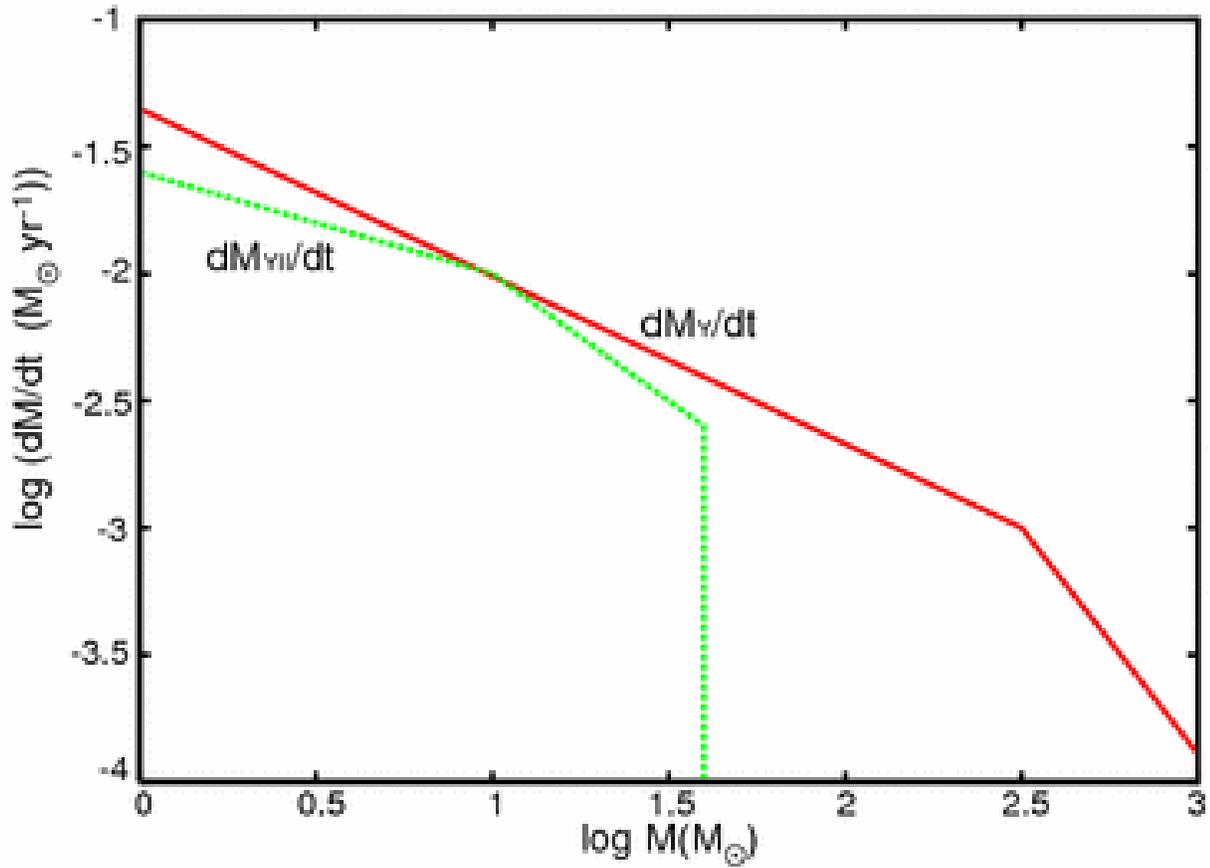}
\caption{Mass accretion rates for the first generation stars
($dM_{\rm Y}/dt$) and the second generation stars ($dM_{\rm
YII}/dt$) as a function of stellar mass $M$ calculated with
cosmological simulations \citep{Yosh06}.
\label{fig:dMYdt}}
\end{figure}

The most important parameter in this work is the mass accretion
rate, $\dot{M}$. We change this value and investigate how the star
evolves and how large the final mass is. We choose the mass
accretion rate calculated with three dimensional cosmological
simulations by \cite{Yosh06} and \cite{Yosh07}. These authors calculated
cosmological structure formation in a ${\Lambda}$CDM universe and
evaluated the accretion rate for Pop III stars. We
approximate $\dot{M}$ from their results as a function of stellar mass
$M$:  \\
\begin{equation}
\dot{M} = \frac{dM_{\rm Y}}{dt} = \left\{
\begin{array}{l}
0.0450 \times M^{-2/3}\, M_{\odot}{\rm yr^{-1}} \\ 
   \qquad   M < 300M_{\odot}  \\
16.3  \times M^{-1.7}\, M_{\odot}{\rm yr^{-1}} \\ 
    \qquad    M \geq 300M_{\odot}
\end{array}
\right.
\label{eqn:MdotY}
\end{equation}
for Pop III.1 stars (by \citealt{Yosh06}), and \\
\begin{equation}
\dot{M} = \frac{dM_{\rm YII}}{dt} = \left\{
\begin{array}{l}
0.0250 \times M^{-0.4}\, M_{\odot}{\rm yr^{-1}} \\ 
       \qquad  M \leq 10M_{\odot}  \\
0.10  \times M^{-1.0}\, M_{\odot}{\rm yr^{-1}} \\ 
       \qquad  10M_{\odot} < M < 40M_{\odot} \\
0 \\ 
       \qquad      M \geq 40M_{\odot}
\end{array}
\right.
\label{eqn:MdotYII}
\end{equation}
for Pop III.2 stars (by \citealt{Yosh07}).
These accretion rates are shown in Figure~\ref{fig:dMYdt}.
We mark the former accretion rate
with a subscript `Y', as $dM_{\rm Y}/dt$ (Equation~\ref{eqn:MdotY}).
We also adopt models with accretion
rates smaller than $dM_{\rm Y}/dt$ by a factor of 2, 3, 10, and 20.
We call these models `Y-series'. The latter
accretion rate (Equation~\ref{eqn:MdotYII}) is labeled `YII', as $dM_{\rm YII}/dt$ (we distinguish this model
from `Y-series'). For this model, the accretion rate is much lower than
$dM_{\rm Y}/dt$, because the gas temperature can become low
in the fossil HII regions around the first generation stars \citep{Yosh07}.
In Table~\ref{tab:ModelsY},
the models we calculate are summarized.

Protostellar feedback effects may considerably affect 
the mass accretion, hence the protostar's growth. Kudritzki (2000) argued that the effect 
of radiation pressure is probably negligible for a Pop III star 
since it has no metal. We discuss more recent studies later in this 
section. \cite{McKe08} 
showed that radiative feedback, if significant, can stop mass 
accretion when the 
stellar mass exceeds $\sim 100M_{\odot}$. If so, Pop III stars around 
this mass range may exist. 

\begin{figure}
\epsscale{1.1}
\plotone{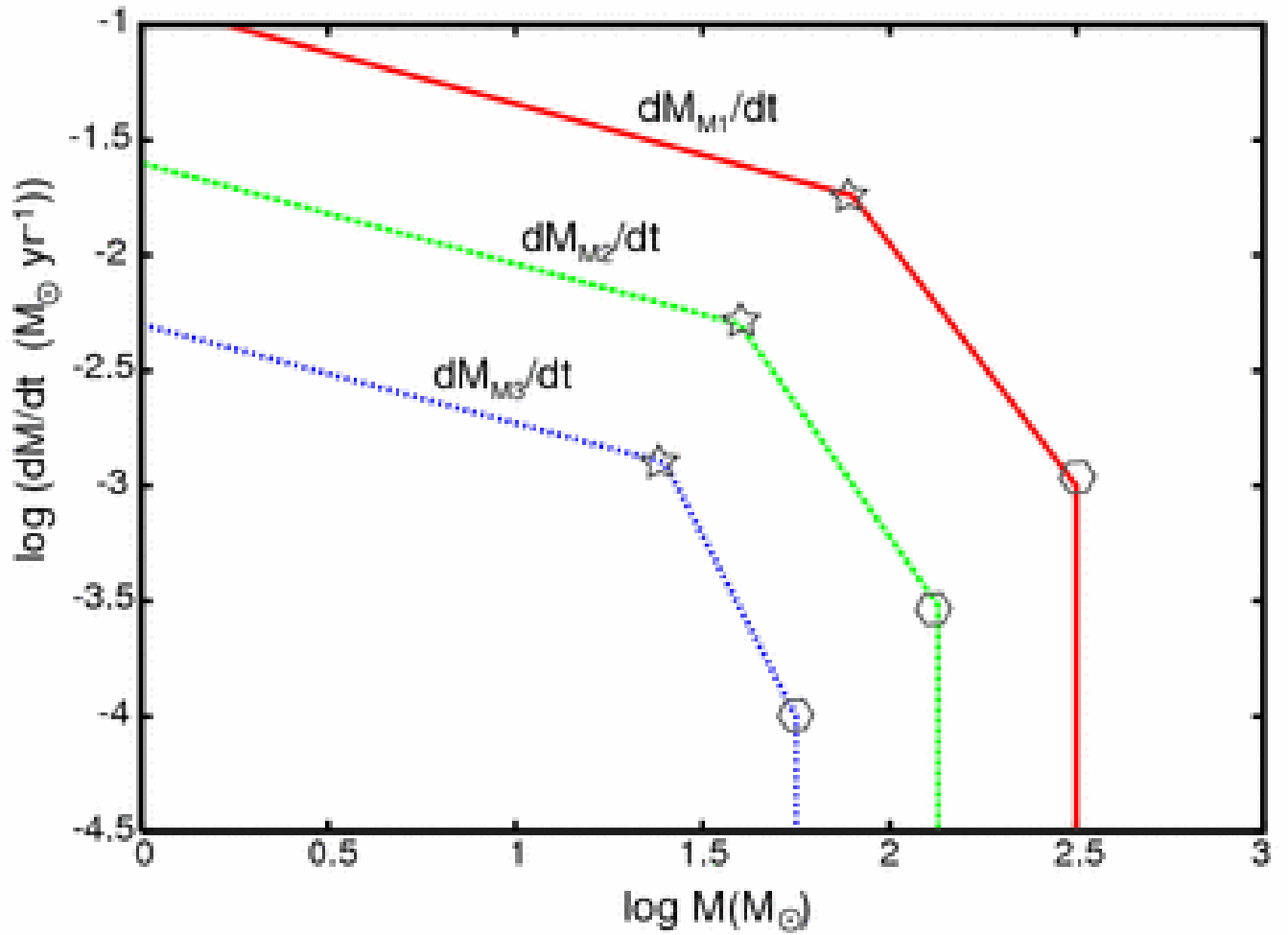}
\caption{Mass accretion rates ($dM_{\rm Mi}/dt$) for the
first generation stars as a
function of stellar mass $M$ by \cite{McKe08}. The open star marks show
the onset of the main-sequence, after which the accretion rate drops
drastically. The open hexagons show the point
where the accretion completely stops.
\label{fig:dMMdt}}
\end{figure}

We also adopt accretion rates by \cite{McKe08}, who take interruption by 
feedback into consideration. In their calculation, the accretion rate is 
very high in the proto-star phase, but when they reach the main-sequence, the accretion 
rate drops drastically owing to evapolation of the accretion disk 
by ionizing photons. During the main-sequence, accretion 
completely stops when outflow and inflow become the same. 
They parametrized a measure of entropy of 
the accreting gas, and subsequent accretion rate varies by more than 
one order of magnitude by changing this parameter. In this paper, 
we adopt typical three accretion rates. We approximate $\dot{M}$ 
from figure 9 in \cite{McKe08} as follows \\
\begin{equation}
\dot{M} = \frac{dM_{\rm M1}}{dt} = \left\{
\begin{array}{l}
0.125 \times M^{-0.44}\, M_{\odot}{\rm yr^{-1}} \\ 
    \qquad   M \leq 81M_{\odot}  \\
178  \times M^{-2.1}\, M_{\odot}{\rm yr^{-1}} \\ 
    \qquad  81M_{\odot} < M < 321M_{\odot} \\
0 \\ 
    \qquad  M \geq 321M_{\odot} ,
\end{array}
\right.
\label{eqn:MdotM1}
\end{equation}
  \\
\begin{equation}
\dot{M} = \frac{dM_{\rm M2}}{dt} = \left\{
\begin{array}{l}
0.0250 \times M^{-0.44}\, M_{\odot}{\rm yr^{-1}} \\ 
    \qquad  M \leq 41M_{\odot}  \\
20.0  \times M^{-2.3}\, M_{\odot}{\rm yr^{-1}} \\ 
    \qquad  41M_{\odot} < M < 135M_{\odot} \\
0 \\ 
    \qquad  M \geq 135M_{\odot} ,
\end{array}
\right.
\label{eqn:MdotM2}
\end{equation}
and \\
\begin{equation}
\dot{M} = \frac{dM_{\rm M3}}{dt} = \left\{
\begin{array}{l}
0.005 \times M^{-0.44}\, M_{\odot}{\rm yr^{-1}} \\ 
    \qquad  M \leq 25M_{\odot}  \\
32  \times M^{-3.2}\, M_{\odot}{\rm yr^{-1}} \\ 
    \qquad 25M_{\odot} < M < 57M_{\odot} \\
0 \\ 
    \qquad M \geq 57M_{\odot}.
\end{array}
\right.
\label{eqn:MdotM3}
\end{equation}
These are marked `M', as $dM_{\rm Mi}/dt$ ($i =1, 2, 3$).
We call these models `M-series' and we summarize them in
Table~\ref{tab:ModelsY}. These accretion rates are shown
in Figure~\ref{fig:dMMdt}.

A Pop III star has initially no metal content by definition, and
thus the line-driven stellar wind is thought to be negligible
\citep{Kudr00}. There are some mechanisms which lead to mass loss
- pulsational instability by the $\epsilon$ mechanism. However,
the time scale of amplification of such oscillations is longer
than the nuclear burning time scale of the main sequence ($\sim
10^6$ years) for Pop III stars, and so a star may evolve without
losing a significant fraction of its initial mass \citep{Ibra81,
Bara01, Nomo03}. On the other hand, there are other authors who
carried out stellar evolution with mass loss even for very
low-metallicity massive stars. \cite{Meyn06} and \cite{Hirs07}
calculated pre-supernova evolution of very-metal poor stars (down
to $Z = 5 \times 10^{-7} Z_{\odot}$) with rotation and found that
such metal-poor stars can lose mass because the CNO elements which
are synthesized in the deep interior are transported to the
surface due to the rotational mixing. As long as mass accretion
continues a typical accretion rate ($\gsim 10^{-3} M_{\odot}{\rm
yr}^{-1}$) is much higher than the mass loss rate, and hence in
the present study mass loss is not considered.


\section{RESULTS}\label{sec:Evolution}

\subsection{Pre Main-Sequence Phase}\label{sec:PreMain}

Our evolutionary calculation starts from $M = 1.5M_{\odot}$.
Figure~\ref{fig:MRrelations} shows the evolutionary change in the
stellar radius $R$ as $M$ increases for each model. It shows a few
distinct evolutionary phases: pre main-sequence phase,
main-sequence phase, and later phase. Each phase is partitioned
with marks.

Because of the gravitational energy release and associated heating,
the stellar radius quickly increases to as large as
$R \sim 10^2 R_{\odot}$. At this point,
$\dot{M} \sim 10^{-2} M_{\odot}{\rm yr}^{-1}$.
After this rapid expansion,
the star settles into the stage where the radius increases only
gradually while adjusting to the mass accretion rate (see also
Figure 3 in \citealt{Neo76}). At this stage the
stellar radius is given approximately by the stellar mass and mass accretion rate as: \\
\begin{equation}
R \propto M^{0.27}{\dot M}^{0.41},
\label{eqn:RMMdot}
\end{equation}
which is similar to the results by
\cite{Stah86} and \cite{Omuk03}. In the subsequent evolution,
our models exhibit the similar trend to those shown in
Omukai \& Palla (2003; Fig.1) and Yoshida et al. (2006; Fig.13).

We started our calculations with mass accretion at $M = 1.5M_{\odot}$. In
more realistic calculations \cite{Omuk03} and \cite{Yosh06}
started their simulation of proto-stellar evolution with mass
accretion with the very first core mass of $M \sim 10^{-3}M_{\odot}$,
although their calculation ends at the end of the proto-star
phase. The time it takes for the star to grow from
$10^{-3}M_{\odot}$ to $1.5M_{\odot}$ is negligible compared with
the lifetime of later phases, and this initial stage does not affect
how massive the star can grow.

After the proto-star phase, the evolution depends on the timescales of
accretion, $\tau_{\rm acc} = M/{\dot M}$, relative to that of
Kelvin-Helmholz (KH) contraction, $\tau_{\rm KH}$, as shown in
Figure~\ref{fig:tauMtauKH}.  \\
(1) During the stage with the filled square in
Figures~\ref{fig:MRrelations} and ~\ref{fig:tauMtauKH},
$\tau_{\rm acc} \sim \tau_{\rm KH}$, so that the radiative luminosity
($L$) is supplied by the gravitational energy release due to mass
accretion. The star does not contract, i.e., the radius stays almost
constant.  \\
(2) During the stage between the filled square and the star mark
in Figures~\ref{fig:MRrelations} and ~\ref{fig:tauMtauKH}, the
accretion timescale gets longer than the radiative energy loss,
i.e., $\tau_{\rm acc} > \tau_{\rm KH}$. Thus the KH contraction
releases the gravitational energy at enough rates to supply the
radiative luminosity. Consequently the star contracts as seen as
the decrease in the radius (Figure~\ref{fig:MRrelations}). During
these stages our results reproduce a similar trend as found by
\cite{Yosh06}.

\subsection{Main-Sequence Phase}\label{sec:MainSequence}

\begin{figure}
\epsscale{.80}
\plotone{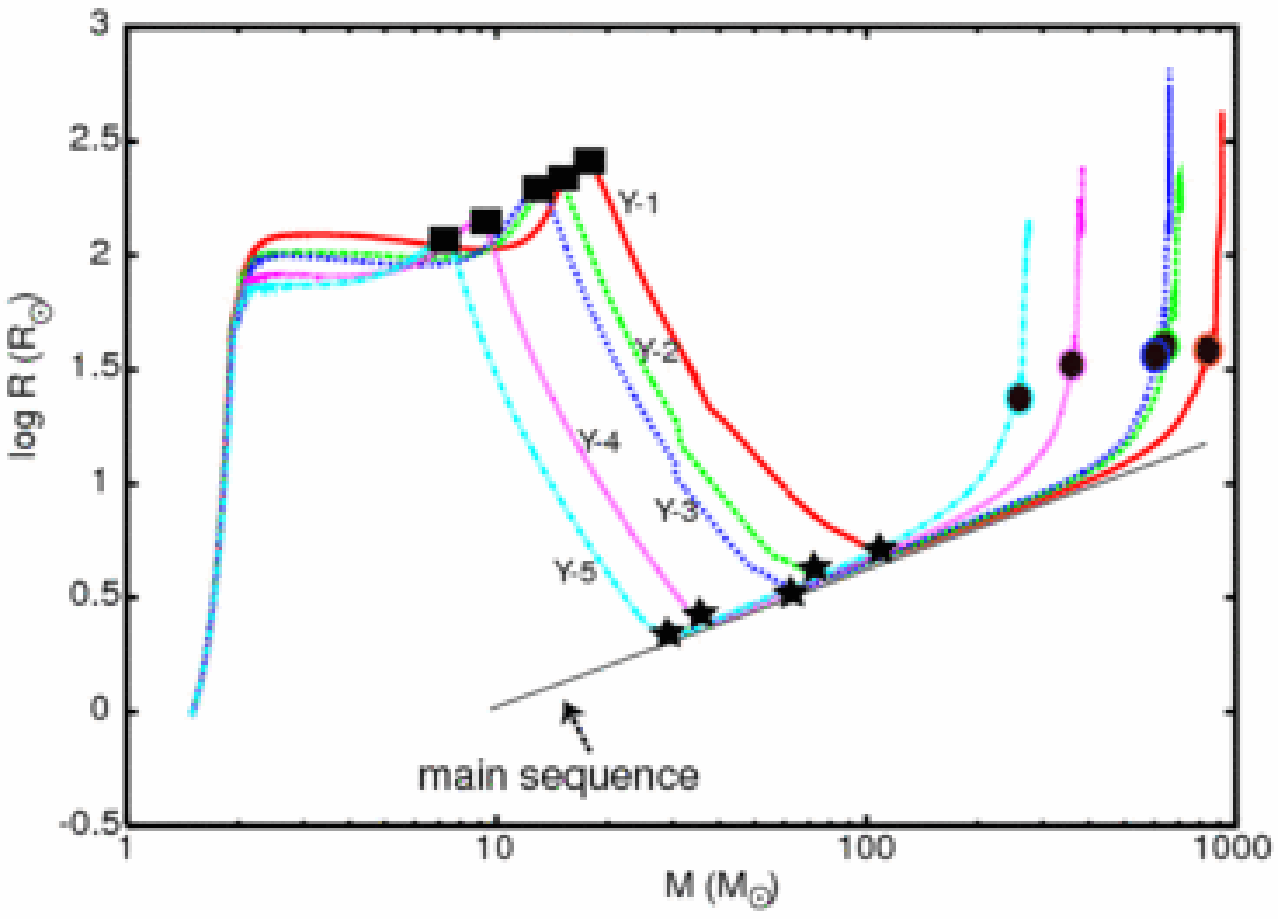}
\plotone{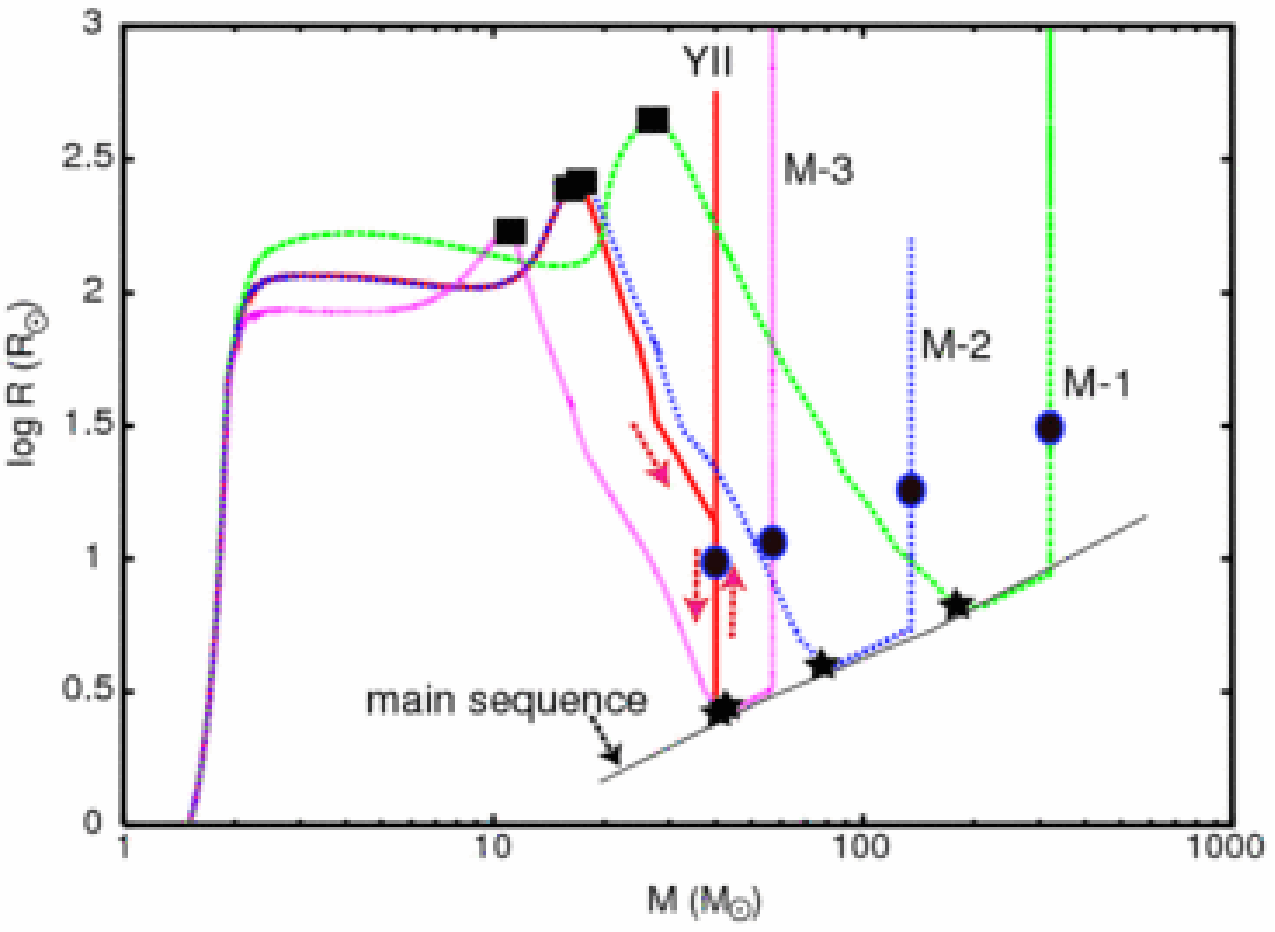}
\caption{Changes in stellar radius, $R$, for stars evolving with increasing
mass $M$. The top panel shows `Y-series', and the
bottom panel shows model YII and `M-series'. The arrows along the
line of model YII indicates the direction of evolution.
The filled squares on the lines show the beginning of
Kelvin-Helmholtz (KH)
contraction phase (the end of rapid accretion phase). The filled star
marks
are the beginning of the main sequence.
The filled circles show the end of main sequence.
\label{fig:MRrelations}}
\end{figure}

\begin{figure}
\epsscale{.80}
\plotone{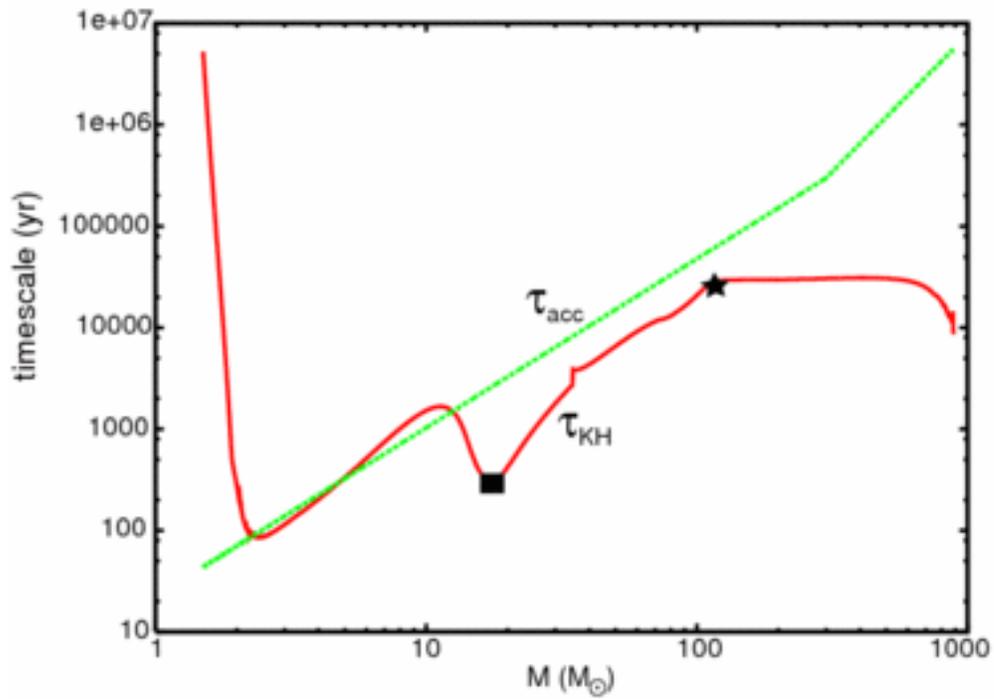}
\caption{The evolutionary change in the timescales of
accretion, $\tau_{\rm acc} = M/{\dot M}$, and Kelvin-Helmholz (KH)
contraction, $\tau_{\rm KH}$, for model Y-1 with increasing $M$.
\label{fig:tauMtauKH}}
\end{figure}

In the stellar interior the temperature rises during the KH contraction.
Because of the absence of primordial CNO elements, hydrogen burning takes place
through the $pp$-chain, which does not produce high enough
nuclear energy generation rate $L_{\rm n}$
to stop the KH contraction (i.e., $L_{\rm n} < L$).
Eventually, the central temperature ($T_c$) exceeds
$\sim 10^8$ K and the $3\alpha$ reaction starts to produce
$^{12}{\rm C}$. When the mass fraction of $^{12}{\rm C}$ reaches
$X(^{12}{\rm C}) \sim 10^{-10}$ at $T_c \sim 10^8$ K, the CNO cycle
starts to produce high enough $L_{\rm n}$. The star then
settles into the main-sequence phase with $L \sim L_{\rm n}$.
This is approximately the turning point in
the stellar radius after the KH contraction (see
Figure~\ref{fig:MRrelations}). The starting point of the main-sequence is 
indicated by the filled star marks in Figure~\ref{fig:MRrelations}. 

For model Y-1, it takes $\sim 10^5$ yr for the star to reach the 
main-sequence, and at this stage the stellar mass reaches several 
$\times 10 - 100M_{\odot}$, already in massive star range. Model 
YII is the only case in which accretion stops during the KH 
contraction phase. For all other models, accretion continues after 
reaching the main-sequence . Pop III massive stars with $M \gsim 
10M_{\odot}$ keep the central temperature as high as $T_c \sim 
1.0-1.5 \times 10^8$ K through the main-sequence. The stellar 
luminosity is of the order of $\sim 10^6 -10^7 L_{\odot}$. The 
main source of nuclear energy generation is the CNO cycle, and 
such high temerature is necessary to supply high enough nuclear 
luminosity to keep $L \sim L_{\rm n}$ with $X({\rm CNO}) \sim 
10^{-10}-10^{-8}$. For `M-series', accretion stops during the 
main-sequence. For YII model, accretion stops before the 
main-sequence. After the accretion stops, the star evolves 
vertically (i.e., $M =$ constant) in Figure~\ref{fig:MRrelations}. 
For `Y-series', mass accretion continues throughout the evolution. 
The accretion rate has decreased to ${\tau}_{\rm acc} \gg 
\tau_{\rm KH}$, so that the star changes its structure to adjust 
to the main-sequence structure of the increased $M$. Thus the star 
evolves approximately along the main-sequence line in 
Figure~\ref{fig:MRrelations} where the stellar radius increases 
with mass ($R \propto M^{0.75}$). The main-sequence phase is the longest in the stellar 
life, and thus the final mass is mostly determined by how much 
mass accretes through the main sequence. The filled circles in 
Figure~\ref{fig:MRrelations} show the end of the main sequence 
(hydrogen exhauston at the center). The star grows to be 
very-massive. The details are sammerized in 
section~\ref{sec:Lifetime}. After central hydrogen burning, 
central helium burning follows. The star expands and the stellar 
radius becomes more than 10 times larger at the end of helium 
burning than that on the main sequence. 
 
\begin{figure}
\epsscale{.80}
\plotone{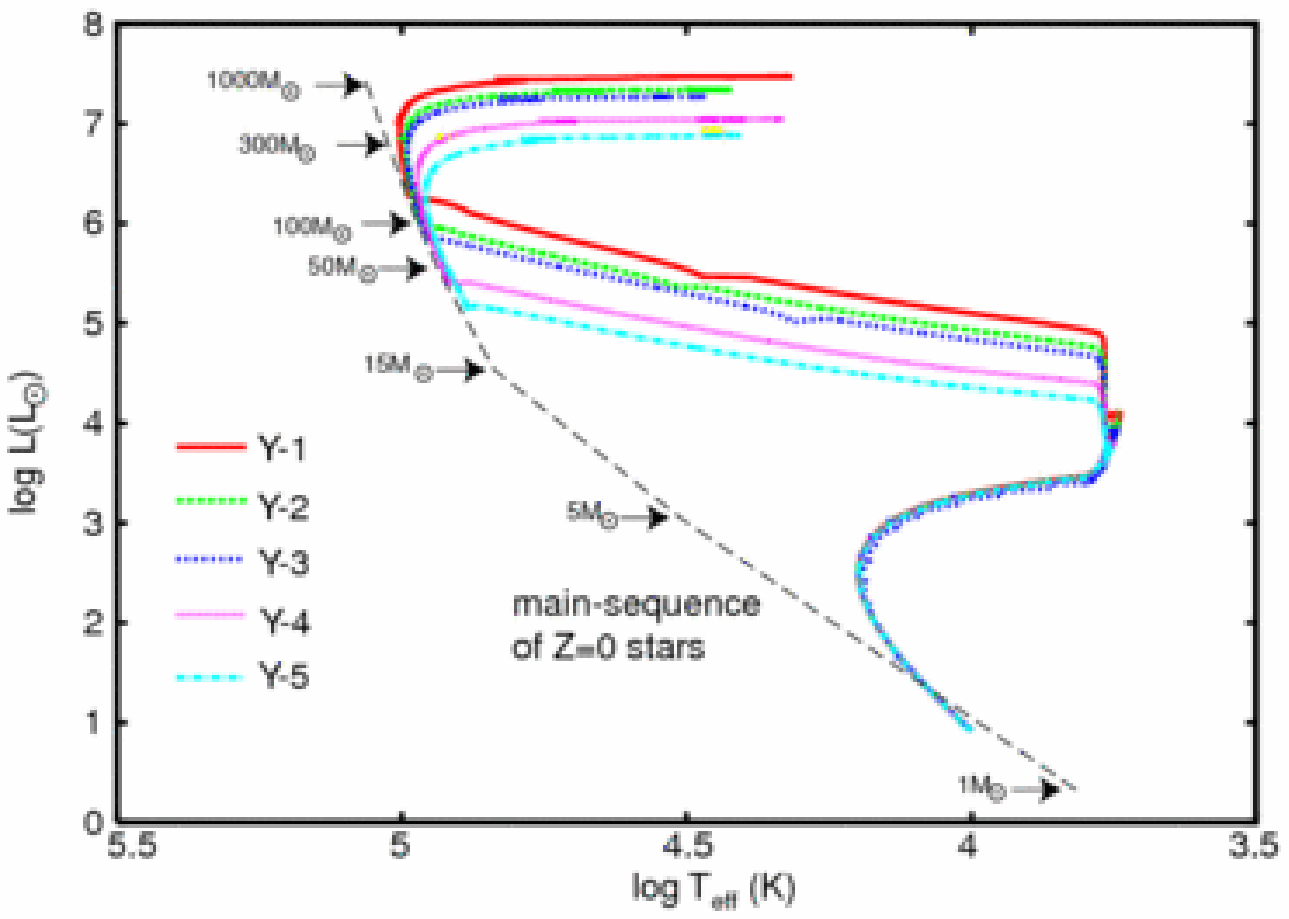}
\plotone{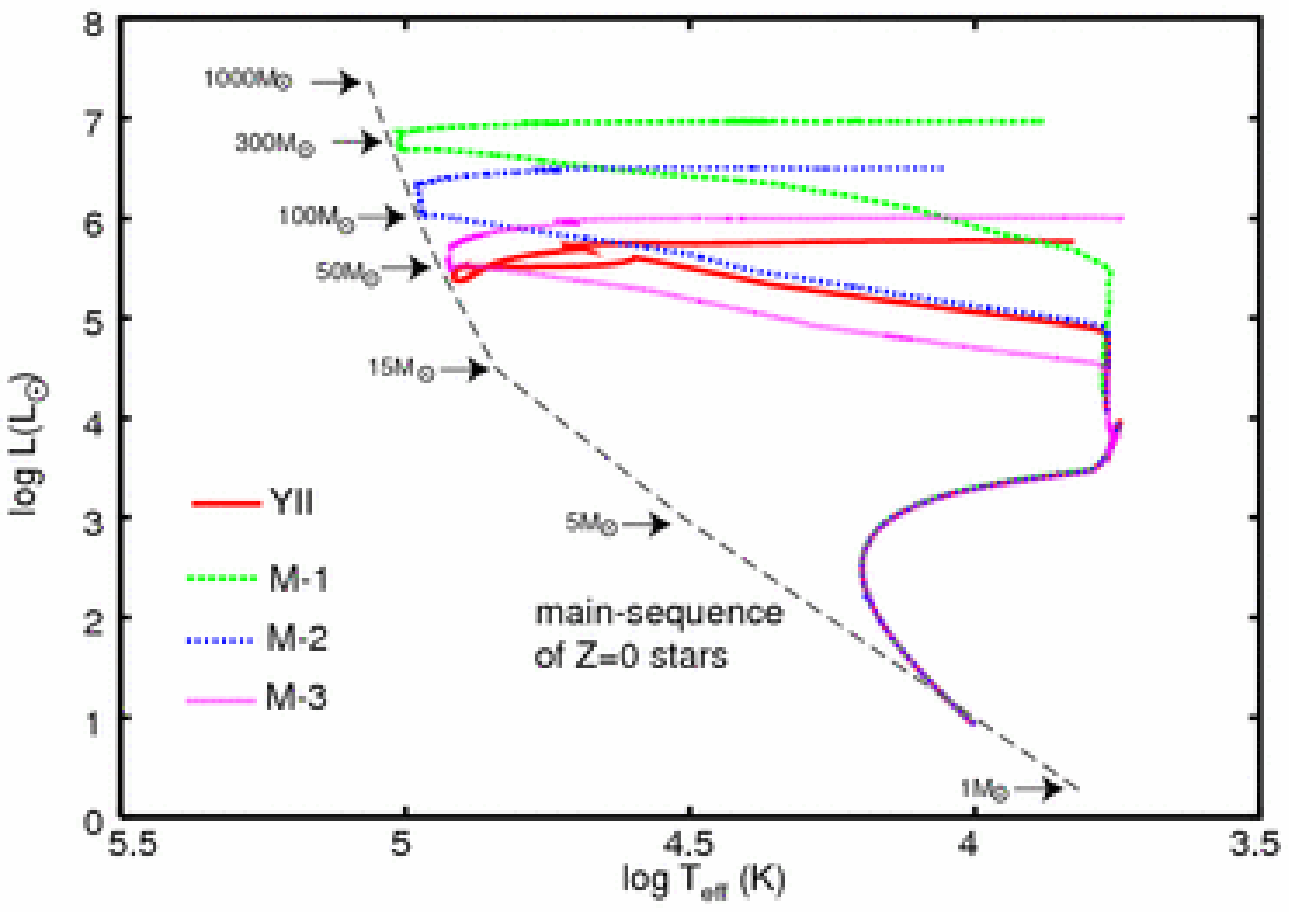}
\caption{The evolution of stars with mass accretion on the HR diagram.
Top panel shows `Y-series', and
bottom panel shows model YII and `M-series'.
The long dashed line shows the location of main sequence stars
with $Z=0$. We adopt the location for $M \leq 100M_{\odot}$ stars from \cite{Mari01}, and for $M \geq 100M_{\odot}$ stars from \cite{Ohku06}.
\label{fig:HRevolve}}
\end{figure}

Figure~\ref{fig:HRevolve} shows the evolution of each model on the
HR diagram. These figures correspond to Figure 8 in \cite{Omuk03}.
During the early mass accretion phase, the stellar radius gets
larger and the surface effective temperature decreases to as low
as $\sim 6000$K. The ${\rm H}^-$ bound-free opacity is the main
source of opacity around this temperature. After this phase, the
star undergoes the KH contraction to decrease its radius. The
effective temperature gets higher to reach $\sim 10^5$ K. The main
opacity source changes from ${\rm H}^-$ bound-free absorption to
electron scattering at such high temperatures. The star reaches
the main-sequence and evolves on the HR diagram along the
main-sequence line with increasing mass. As hydrogen is consumed,
the star leaves the the main-sequence to decrease its effective
temperature. At the end of hydrogen burning, the stellar mass
reaches near its final mass.

\begin{figure}
\epsscale{.80}
\plotone{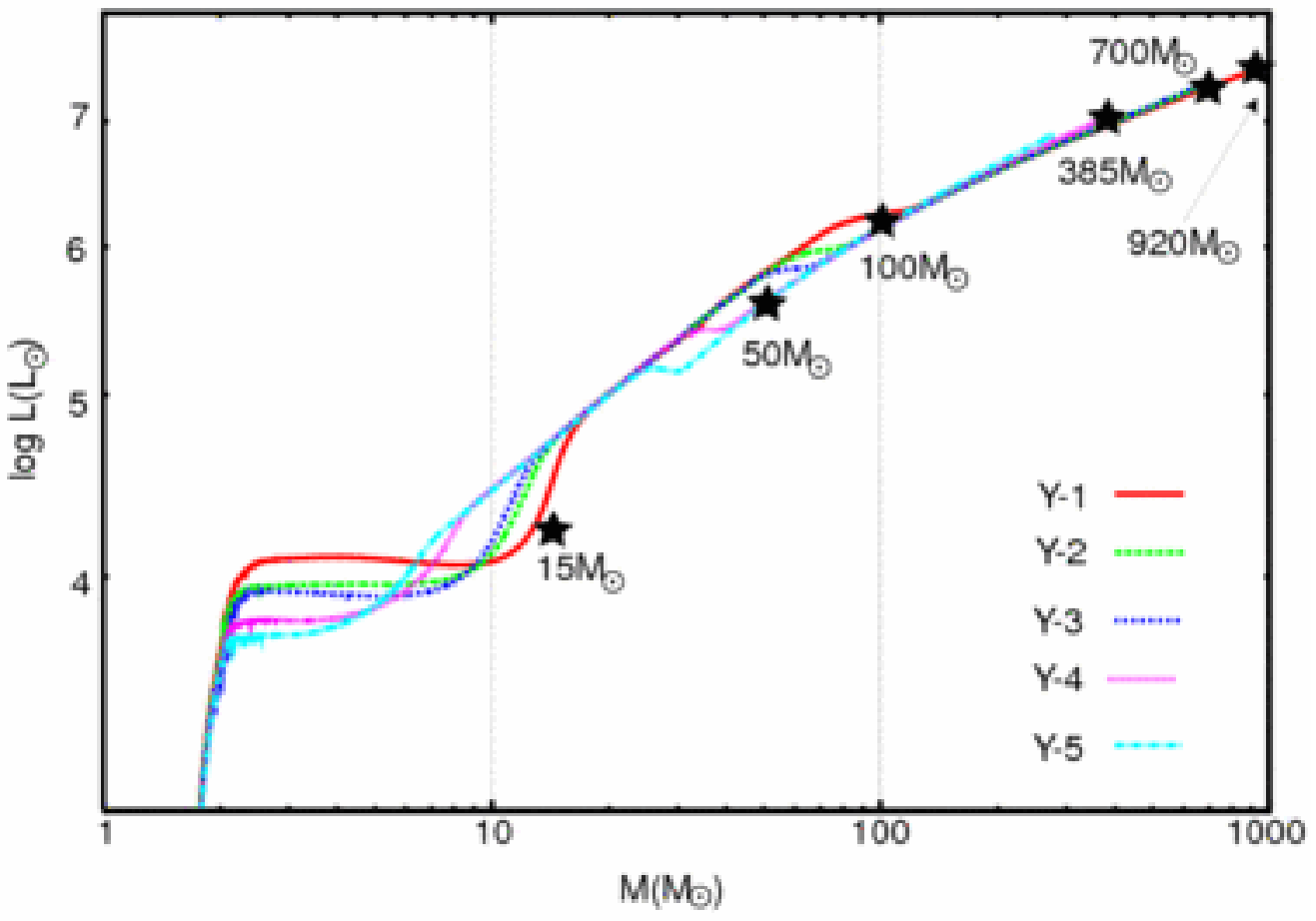}
\plotone{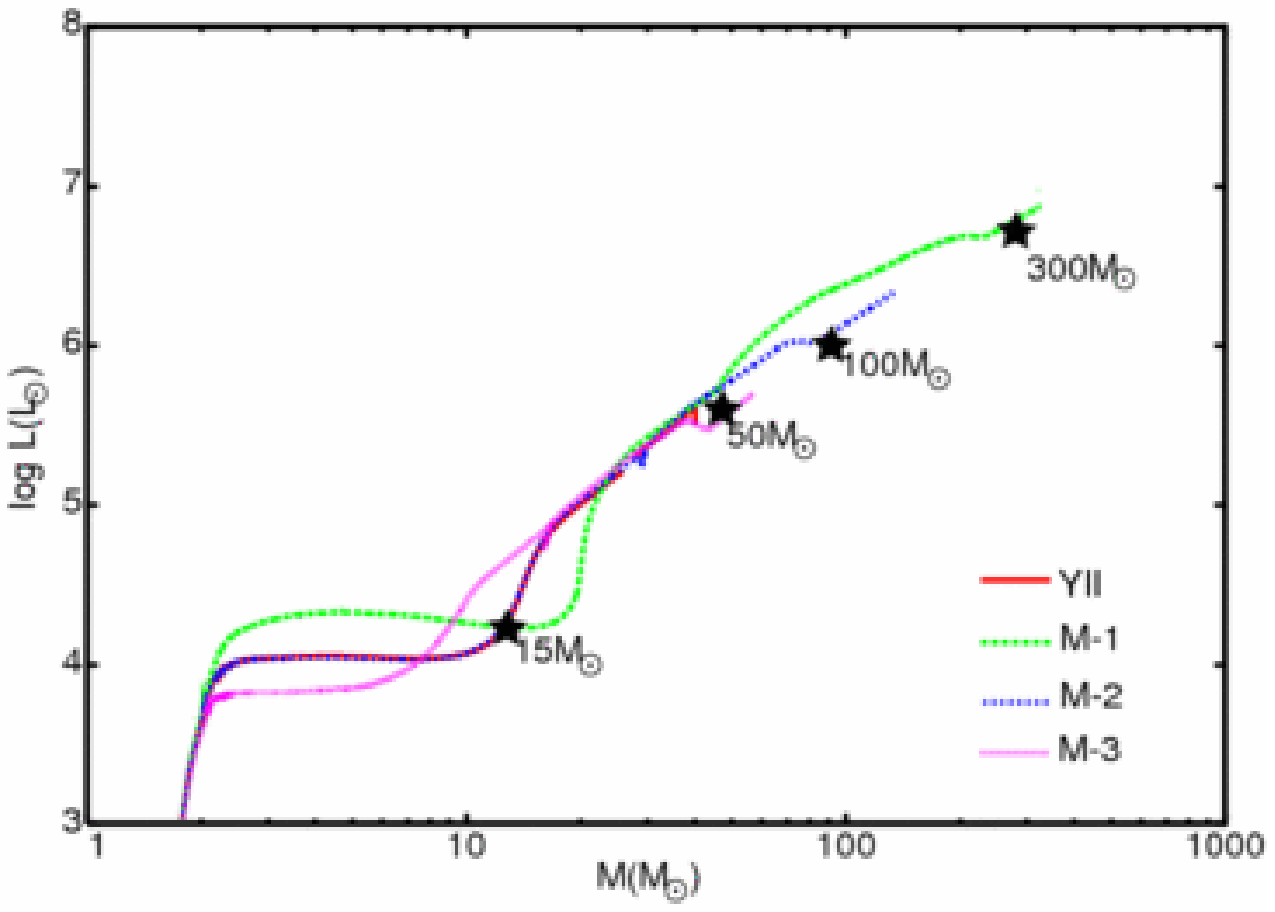}
\caption{The evolutionary change in the stellar luminosity with the
increasing mass, $M$, for `Y-series' (top) and
model YII and `M-series' (bottom).
The filled star marks indicate the main-sequence of
stars without mass accretion.
\label{fig:MLrelations}}
\end{figure}

Figure~\ref{fig:MLrelations} shows the evolutionary changes in stellar
lunminosity with increasing mass.
For comparison, the main-sequence points of several stars with no mass
accretion are also shown.
After reaching the main-sequence, the stars evolve along the
main-sequence lines of corresponding stars with no mass accretion.
Figure~\ref{fig:RTtrack} shows the evolutionary tracks of the
central density ($\rho_c$) and central temperature at later phases
through carbon, neon, oxygen, and silicon burning for four models.
For comparison, we also show those of the 25$M_{\odot}$ star
(\citealt{Umed02}) and the 1000$M_{\odot}$ star (\citealt{Ohku06})
with no mass accretion. Generally, more massive stars have higher
entropies (lower densities) at the same temperatures. For the
model with mass accretion, the central entropy is determined when
the stellar mass is small, and remains low because of the small
${\tau}_{\rm acc}/{\tau}_{\rm KH}$ even when the stellar mass
becomes much larger. After H is ignited, the central
density-temperature path moves toward higher entropy, while the
central temperature is kept at $T_c \sim 1 \times 10^8$ K.

\subsection{Stellar Lifetime and the Final Mass}\label{sec:Lifetime}

After helium burning (at log $T_c$ (K) $\sim 8.5$), the star with
mass accretion evolves almost parallel to stars with no mass
accretion in the ($\rho_c, T_c$) diagram. This is because the
evolutionary timescale is $\lsim 10^4$ yr from central helium
exhaustion to core-collapse or explosion, and $\lsim 1$ yr after
$T_c$ reaches $10^9$K which is much shorter than the accretion
timescale (${\tau}_{\rm acc} = M/{\dot M}$). Thus, the star
evolves with negligible effects of mass accretion, and the stellar
mass can be regarded as constant, reaching its final value
$M=M_{\rm f}$.

The final mass, $M_{\rm f}$, is determined by the stellar lifetime
and mass accretion rate. In Table~\ref{tab:fates} these results
are summarized. The final mass for `Y-series' is very large. This
is primarily due to the long lifetime of the star. \cite{Ohku06}
calculated the evolution of CVMS models with $M = 500M_{\odot}$
and $1000M_{\odot}$ with no accretion. The lifetime of CVMS ($M
\gsim 300M_{\odot}$) without accretion is $\sim 2 \times 10^6$ yr,
which does not sensitively depend on the mass because the $M-L$
relation is close to $L \propto M$. On the other hand, our models
with mass accretion have longer lifetime, $\gsim 3 \times 10^6$
yr, even if the final mass exceeds $300M_{\odot}$. Stars with
mass accretion evolve through the lower mass phase. This makes
stellar lifetimes longer. On the other hand, for model M-1, the
lifetime is as short as $2.2 \times 10^6$ yr, much shorter than
models Y-4 and Y-5, even though the final mass is in the same
range. For model M-1, accretion stops after $1 \times 10^5$ yr,
just after hydrogen burning starts. At this stage the star has
already reached the VMS range so that the luminosity (and thus
hydrogen consumption speed) is very high.

\subsection{Final Phases of Evolution and Final Fate}\label{sec:Phases}

 For models Y-1, 2, 3, 4, and M-1,
$M_{\rm f}$ exceeds $300M_{\odot}$ so that they end their life with
iron core-collapse. Model Y-5 has $M_{\rm f} =
275M_{\odot}$ and ends its life as a PISN. For models YII,
M-2, and M-3 the final mass is too small to become a PISN. These
stars collapse to form a black hole. Whether a
star explodes as a PISN or causes iron core collapse is decisively
important for galactic chemical evolution and the origin of
elements. Main features of the final phases of evolution of these 
models are summarized
below and more details will be published elsewhere.

When central helium is exhausted, the CO core contraction accelerates. 
Figure~\ref{fig:TauT2} shows the time scale of the increase in the 
central temperature 
(${\tau}_T \equiv dt/d{\rm ln}\,T_c$) for models Y-1 ($M_{\rm f} = 915M_{\odot}$) and 
YII ($M_{\rm f} = 40M_{\odot}$). 
From log $T_c$ (K) $=8.6$ to 9.2, ${\tau}_T$ decreases 
drastically from $\sim 10^5$ yr to $\sim 10^{-4}$ yr for model Y-1. 
At log $T_c$ (K) $= 9.2$, model Y-1 is dynamically collapsing with such short 
$\tau_{T}$, while less massive model YII still undergoes quasi-static evolution with $\tau_{T} = 10^0$ yr.  For all cases, the core evolution accelerates, being decoupled from the envelope.

\subsubsection{Stars with $M_{\rm f} \simeq 140 - 300M_{\odot}$}
Model Y-5 ends as a PISN because it enters
the pair-instability region at log $T_c$ (K) $\sim 9$ as seen in
Figure~\ref{fig:RTtrack}, where the dynamically unstable region for the 
core with the adiabatic exponent $\gamma < 4/3$ is indicated. The
central region collapses, which induces thermonuclear runaway of
oxygen and silicon burning due to the rapid temperature increase.
The thermal pressure generated by the runaway nuclear burning stops the
core contraction and expands the star
(\citealt{Raka67, Ober83, Bond84, Glat85, Woos86}).
This appears as the turning point of the evolutionary tracks in
Figure~\ref{fig:RTtrack}. The calculation is stopped when the total 
energy of the star becomes positive, which leads to the total disruption 
of the star. 

\subsubsection{Stars with $M_{\rm f} \gsim 300M_{\odot}$}
Models Y-1, 2, 3, and 4 enter
the pair-instability region in Figure~\ref{fig:RTtrack}. However, they do not explode
as PISNe but undergo core-collapse to become
black holes.
These very massive stars have such large gravitational binding energy that
the nuclear energy release can not exceed this binding energy.
More details of the evolutionary processes of model Y-1 are as follows.

Figure~\ref{fig:Kippen} shows the chemical
evolution in the interior of model Y-1 ($M_{\rm f} = 915M_{\odot}$). 
Helium is exhausted in the core 
around $t_{\rm collapse} - t \sim 10^4$ yr.
When carbon is ignited at $t_{\rm collapse} - t \sim 10^{-4}$ yr, the central
region enters the pair instability region. 

During collapse ($t_{\rm collapse} - t \lsim 10^{-6}$ yr), the
silicon layer and iron core rapidly grow in mass. This is because
in the intermediate region ($M_{\rm r} \sim 100 -200
M_{\odot}$) oxygen and silicon burn explosively with the time
scale of core collapse. 
Figure~\ref{fig:Snap4-Y1} shows the abundance distribution for
model Y-1 at log $T_c$
(K) $= 10.3$ and log ${\rho}_c$ (${\rm g\;cm^{-3}}$) $= 10.0$.
One can see that the mass of the iron core of CVMS is large ($\sim
20 -25 \%$ of the total stellar mass), much larger than ordinary
massive stars such as $\sim 10$ \% in the 25$M_\odot$ star. This
result agrees with that of 500 and 1000$M_{\odot}$ stars in
\cite{Ohku06}. 


\subsubsection{Stars with $M_{\rm f} \lsim 140M_{\odot}$}

\begin{figure}
\epsscale{.80}
\plotone{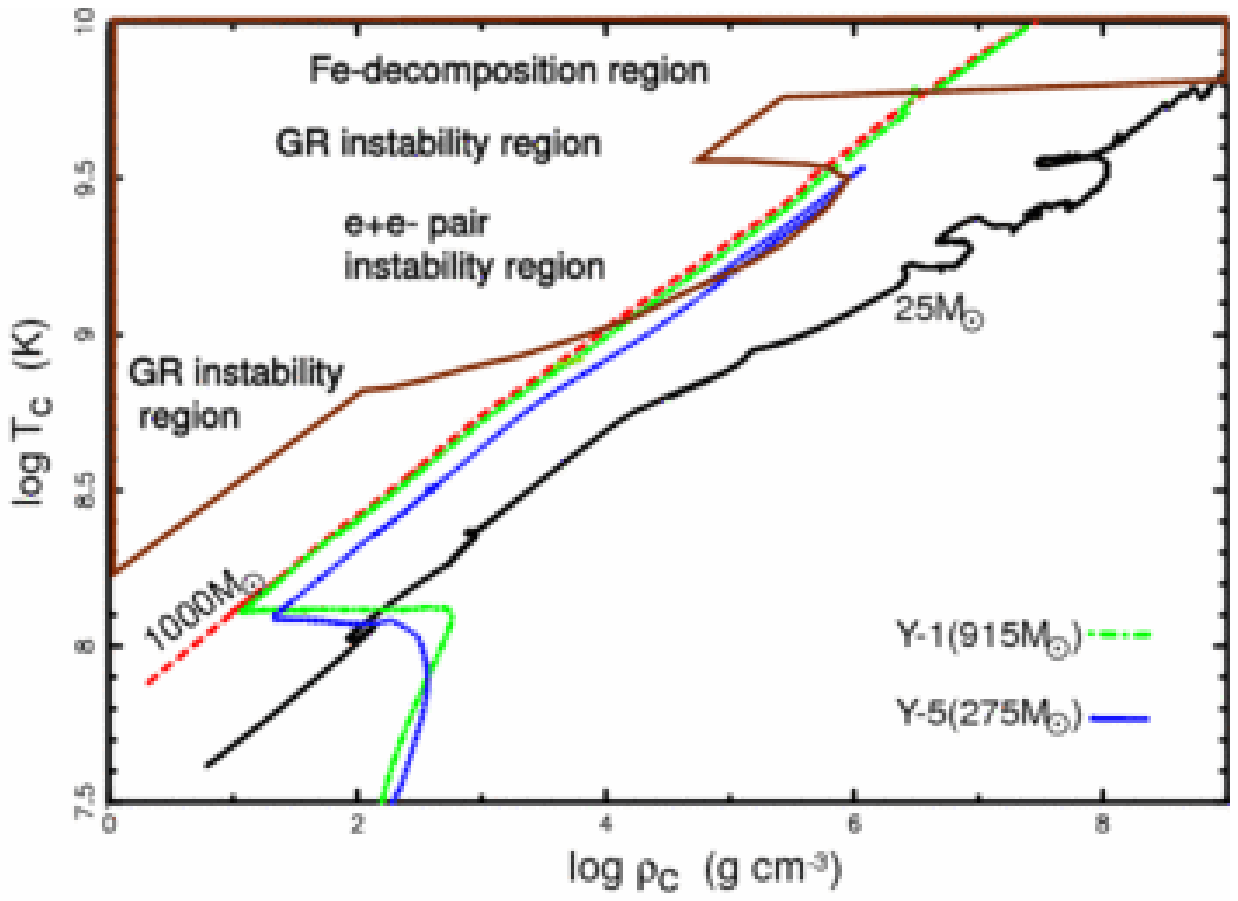}
\plotone{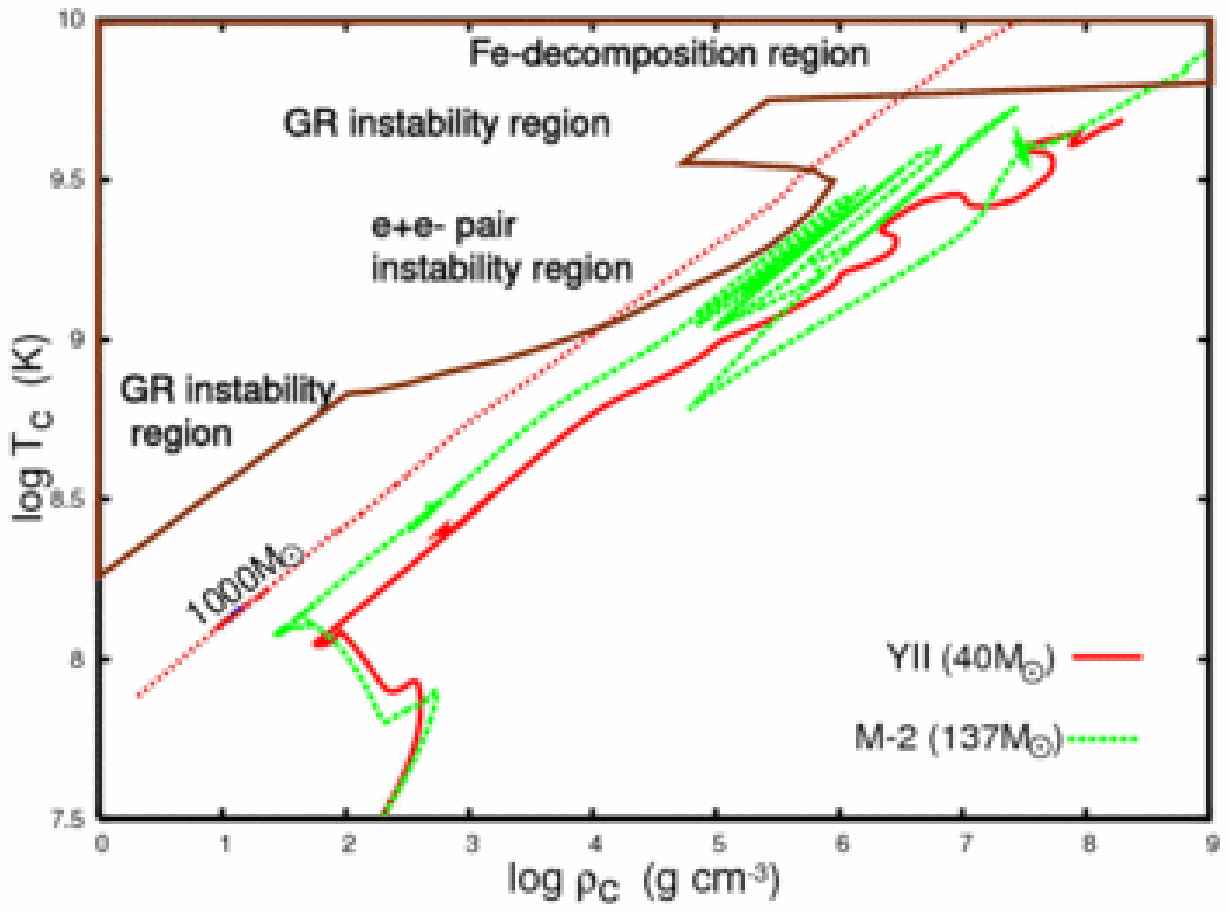}
\caption{The evolutionary tracks of the central temperature ($T_c$) and central
density (${\rho}_c$) of models Y-1 and Y-5 (top panel) and models YII
and M-2 (bottom panel).
The figures in the brackets are the final mass $M_{\rm f}$.
For comparison, the tracks of 25$M_{\odot}$
(\citealt{Umed02}) and 1000$M_{\odot}$ (\citealt{Ohku06})
stars are also shown. The dynamically instable regions with the adiabatic 
exponent $\gamma < 4/3$ are indicated. 
\label{fig:RTtrack}}
\end{figure}

\begin{figure}
\epsscale{.80}
\plotone{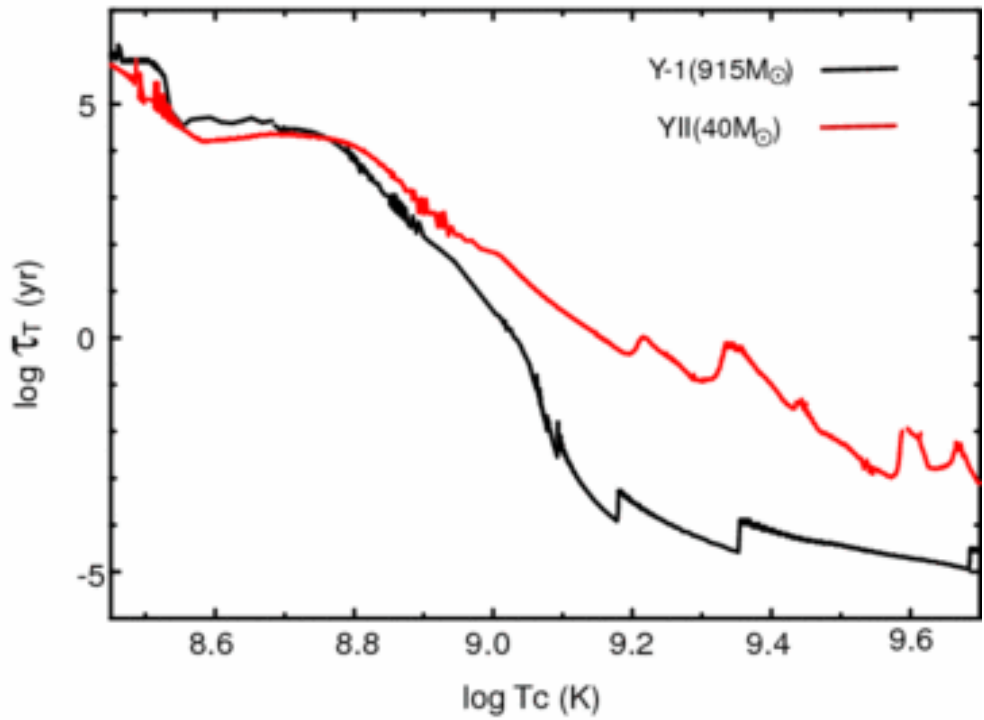}
\caption{Time scale for central temperature increase, $\tau_{\rm T}$, as a function of 
central temperature log $T_c$ after helium burning for models 
Y-1 and YII.    
\label{fig:TauT2}}
\end{figure}

\begin{figure}
\epsscale{1.0}
\plotone{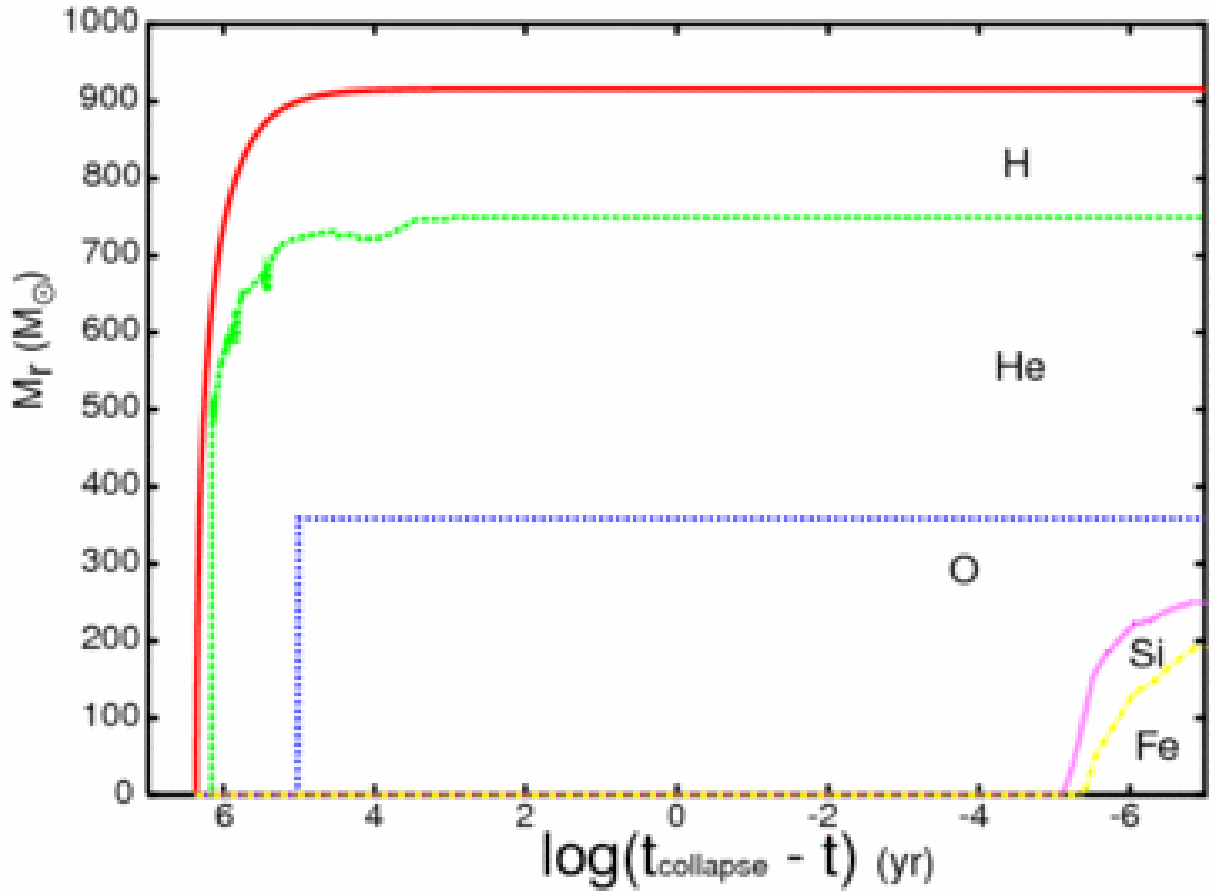}
\caption{Evolution of the chemical structure as a function of time left until the
core collapse for model Y-1 ($M_{\rm f} = 915M_{\odot}$).
In each
layer the indicated element has the largest
mass fraction. \label{fig:Kippen}}
\end{figure}

\begin{figure}
\epsscale{1.0}
\plotone{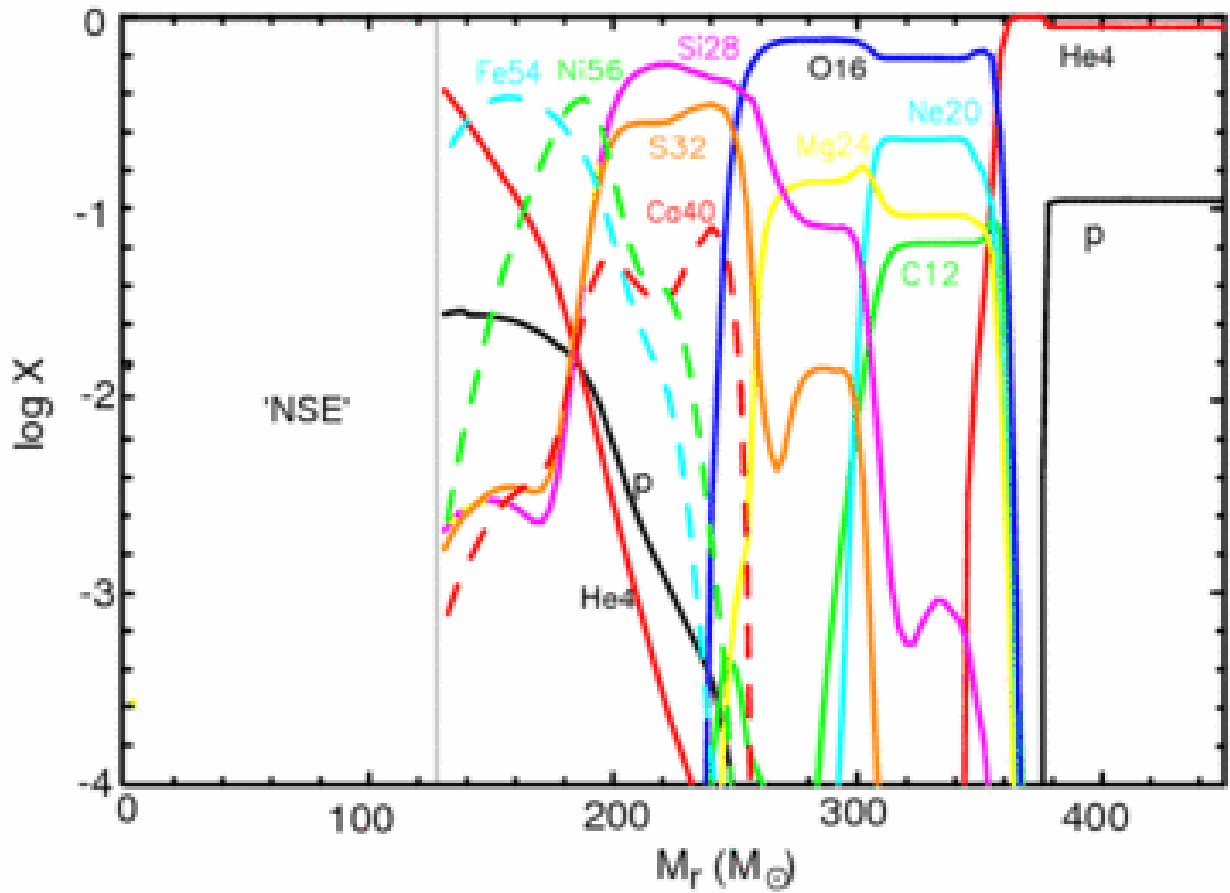}
\caption{Abundance (mass fraction) distribution for model
Y-1 ($M_{\rm f} = 915M_{\odot}$) when log $T_c$ (K) $= 10.3$ and
log $\rho_c ({\rm g\;cm^{-3}}) = 10.0$.
\label{fig:Snap4-Y1}}
\end{figure}

The final mass of model YII ($M_{\rm f} = 40M_{\odot}$) is in the
range of ordinary massive stars ($\lsim 60M_{\odot}$), and so the
evolutionary track is similar to that of 25$M_{\odot}$ in the top
panel of Figure~\ref{fig:RTtrack}. It ends with core-collapse and forms a black
hole.

Model M-2 ($M_{\rm f} = 135M_{\odot}$) undergoes the core oscillation 
as seen in the ($\rho_c, T_c$) evolution 
(Figure~\ref{fig:RTtrack}, lower). The evolutionary path  in the 
($\rho_c, T_c$) plane goes through the marginally stable 
region against the pair-creation instability, so that the adiabatic exponent
$\gamma$ is close enough to induce the pulsational nuclear instability 
due to central oxygen and silicon burning. In stars with
$80M_{\odot} \lsim M_{\rm f} \lsim 140M_{\odot}$, slightly under the
PISN mass range, such a core oscillation appears (\citealt{Woos07, 
Umed08}). The nuclear energy released during the oscillation is not large enough to
disrupt the whole star, so that the core contracts and
central temperature and density rise again. The star finally
collapses to form a black hole after several oscillations 
(Figure~\ref{fig:RTtrack}).

\begin{figure}
\epsscale{1.0}
\plotone{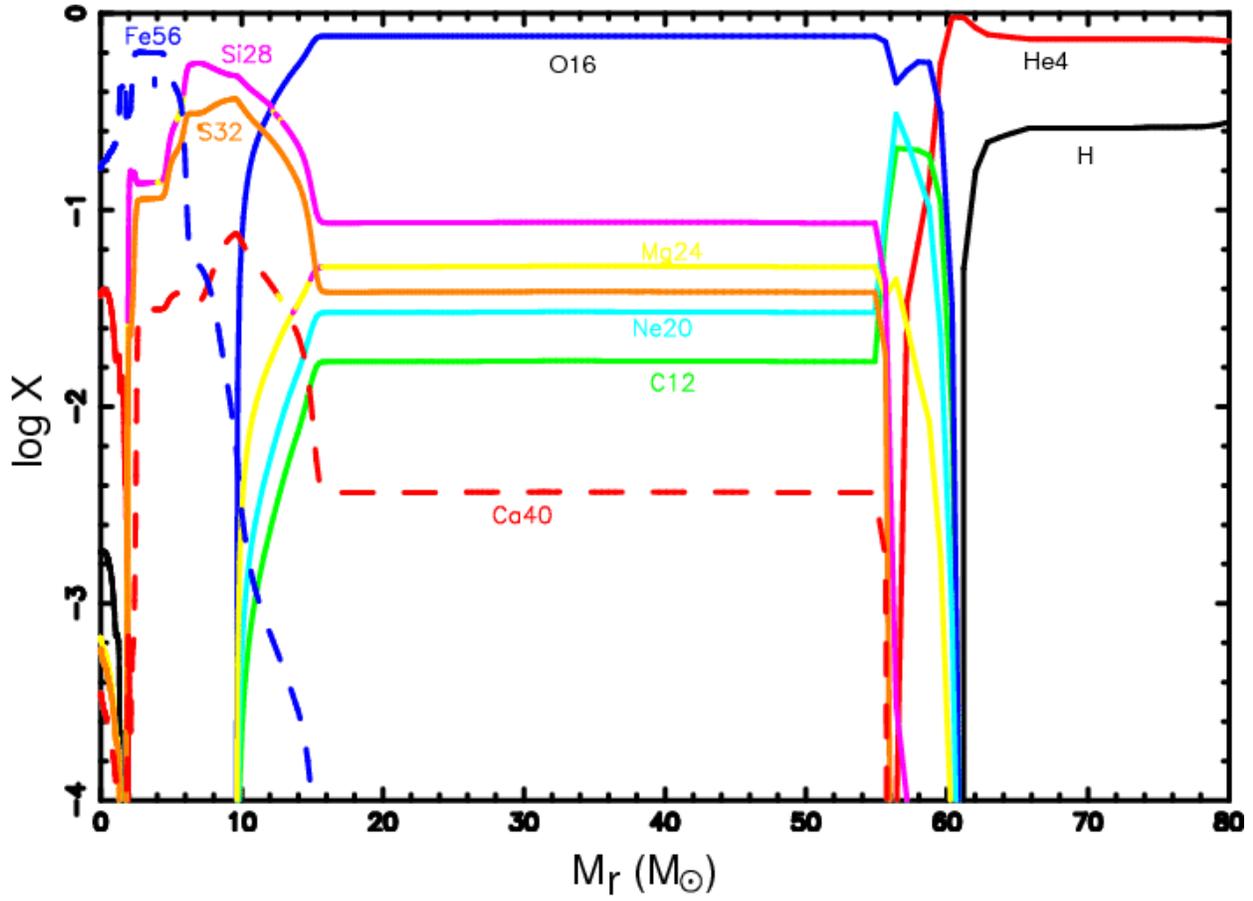}
\caption{Same as Figure~\ref{fig:Snap4-Y1} but for model M-2
($M_{\rm f} = 135M_{\odot}$) when log $T_c$ (K) $ = 9.95$ and
log $\rho_c ({\rm g\;cm^{-3}}) = 9.3$.
\label{fig:Snap4-M2}}
\end{figure}

Figure~\ref{fig:Snap4-M2} shows the abunance distribution for
model M-2 ($M_{\rm f} = 135M_{\odot}$) at 
log $T_c$ (K) $= 9.95$, and log ${\rho}_c$
(${\rm g\;cm^{-3}}$) $= 9.3$. Contrary to model Y-1, the iron
core mass is as small as 6$M_\odot$, being 5\% of the stellar
mass. One can see from Figure~\ref{fig:RTtrack} that the
evolutionary track is similar to that of the 25$M_{\odot}$ star at
the collapse stage. The silicon or oxygen layer does not pass through the
pair instability region, unlike the VMS star.
Neither
the silicon or oxygen layer can follow the central iron core
collapse.

\section{DISCUSSION}\label{sec:Discussion}

In this paper, we have calculated the evolution of Pop III stars that
undergo mass accretion. The evolution of
Pop III VMS has been studied by many authors through later nuclear burning stages, collapse, and explosion
(e.g., \citealt{Ober83, Bond84, Arne96, Umed02, Hege02, Ohku06}).
These studies assumed that such stars have their large
mass from their starting point of their life and the same mass has
been held through the evolution.


Previous proto-stellar calculations ended at the onset of hydrogen
burning on the main-sequence. Our results show that with mass
accretion the stellar mass continues to grow after the pre main-sequence
phase through the main-sequence and beyond.
Assuming that radiative feedback is not significant in the
majority of cases,
we have confirmed the conclusion of \cite{Ohku06} that there is a
realistic possibility for the existence of CVMSs as Pop III stars
in the early universe. We find that the final mass can be very
large, being in the range where a star collapses to become an IMBH
($M_{\rm f} \sim 300 - 1000M_{\odot}$).

\cite{Ohku06} argued that these CVMSs contribute significantly to
the metal enrichment of the early universe, while preventing the
overabundance, before ending their life quickly before ordinary
core-collapse supernovae and hypernovae became dominant. \cite{Ohku06} 
also found that the abundance of ejected heavy metals in
this CVMS scenario agrees well with the observed chemical
abundance data from the early universe, in contrast to poor
agreement of the abundance patterns of PISNe. This is because the
mass ratio of each element ejected by PISNs cannot reproduce the
abundance patterns of EMP stars, intergalactic medium (IGM), or
intracluster matter (ICM) (\citealt{Umed02, Hege02, Chie02}).
Furthermore, if the majority of the first generation Pop III.1
stars are CVMSs, they will form an IMBH, not explode as a PISN.
Then, the second generation (Pop III.2) stars formed under the
influence of radiation from Pop III.1 stars are less massive than
$\sim 100M_{\odot}$. This scenario may explain why PISN signatures
are not observed in the abundance patterns.

The presence of CVMS in the early universe, if confirmed, has
important implications for progenitors of IMBHs. There is an
interesting suggestion that some IMBHs may have been found
(\citealt{Bart05}). \cite{Mats01}, by using {\sl Chandra},
reported possible identification of a $\gsim 700M_{\odot}$ black
hole in M82 as an ultra-luminous X-ray source (ULX). More such
detections have been reported as ULXs\citep{Colb99, Maki00,
Maki08} and also as an object in the Galactic center
\citep{Hans04}.

As to the formation of SMBHs, there are several scenarios (e.g.,
\citealt{Rees02, Rees03}). SMBHs may be formed directly from
supermassive halos of dark matter (e.g., \citealt{Marc80, Brom03b,
Bege06}). \cite{Mada01} first suggested that mini halos with black
holes with mass $\sim 150M_{\odot}$ at redshift $z \sim 20$ will
merge with each other successively to eventually form SMBHs.
\cite{Ebis01} suggested a scenario where IMBHs grow to become
SMBHs by merging and swallowing of many of these objects in dense
clusters. Several authors studied detectability of gravitational
waves radiated by IMBH binaries at high redshifts, by
observatories such as Laser Interferometer Space Antenna
(LISA)(\citealt{Sesa07, Mici07, Tana08, Plow09}). If CVMSs
actually existed, they could be considered as natural progenitors
of IMBHs. Our results offer a natural scenario for the formation
of seed black holes responsible for a merger tree model. The
existence of CVMS as first stars will be supported if in the
future neutrinos and gravitational waves are detected from
collapsing CVMSs (\citealt{Suwa09}).

Our present work uses a result from a three-dimensional
cosmological simulation by Yoshida et al. (2006). Volonteri et al.
(2003) have shown that the merger tree scenario starting with
black hole seeds of $\simeq 150M_{\odot}$ at $z \sim 20$ is
consistent with the present day SMBH population estimated from
observed quasar luminosity and mass functions. However, any
relevant model must also be consistent with the presence of SMBHs
massive enough (a few $\times~ 10^9M_{\odot}$) to power the bright
high redshift quasars recently discovered by the Sloan Digital Sky
Survey (see, e.g, a recent review by Fan 2006). Various models
have been proposed to explain these high redshift powerful massive
SMBHs (e.g., \citealt{Volo06, Bege06, Tana08, Li07, Li08}). In
some of these models, the seed black holes have to be formed in
hotter (virial temperature $T_{\rm vir} \geq 10^4$ K) and more
massive ($> 10^8M_{\odot}$) DM halos, and the BHs grow by
supercritical accretion, in order to become massive enough by $z
\simeq 6$. \cite{Tana08} further extended these earlier studies
and examined successful scenarios which can be constrained by both
high redshift and local SMBH populations. In some of these models
(e.g., \citealt{Volo06, Tana08}) the seed black holes are of the
order of $100M_{\odot}$ formed from collapse of Pop III VMSs,
while in some others the seeds are black holes of the order of
$10^{4-5} M_{\odot}$ which are formed directly from DM halos
(e.g., \citealt{Bege06, Tana08}). \cite{Tana08} conclude that the
models with seed black holes from both models can be successful
(to be consistent with both $z=6$ and local quasar populations),
but LISA observations can distinguish between them. The
environment where the seed black holes are formed from more
massive and hotter DM halos is quite different from that of
conventional star formation models such as those of Yoshida et al.
(2006). One possible interesting extension of our current work
will be to adopt supercritical accretion for star formation in the
environment of more massive hotter DM halos.

\cite{Tana08} show that in a successful merger tree scenario for
$z=6$ quasar formation without supercritical accretion, if the
seed black holes are $100M_{\odot}$ formed from Pop III stars and
subsequently accrete $> \sim$ 60\% of the time, these holes must
be formed rather early at $z > 30$. We note that this constraint
may be eased appreciably if the seed black holes are larger
instead, such as IMBHs of $\sim 1000M_{\odot}$ from our CVMSs. The
maximum mass for seed black holes to be formed from CVMSs may
still become larger if the environment of hot, massive DM halos
(of $T_{\rm vir} \geq 10^4$ K) is adopted in our future
calculations.


\section{CONCLUDING REMARKS}\label{sec:Concl}

In the present study, we calculate the evolution of Pop III 
stars whose mass grows from the initial mass of $\sim
1M_{\odot}$ by accreting the surrounding gases. Our calculations
cover a whole evolutionary stages from the pre-main sequence, via
various nuclear burning stages, through the final core collapse or
pair-creation instability phases. We calculate models with various
mass accretion rates: (1) stellar mass dependent accretion rates which are
derived from cosmological simulations of early structure formation
(Pop III.1 stars), based on the low mass dark matter halos at
redshifts $z \sim 20$, and (2) mass dependent
accretion rates for zero-metallicity but second generation (Pop
III.2) stars which are affected by radiation from the first
generation (Pop III.1) stars. For comparison, we also adopt mass dependent accretion rates which
are affected by radiative feedback.

The final mass of Pop III.1 stars can be very large 
($M \sim 1000M_{\odot}$), beyond the PISNe mass range 
($140M_{\odot} - 300M_{\odot}$). Such massive stars
form IMBHs, which may be the seeds for the formation 
of supermassive black holes. Intriguingly, it is theoretically 
suggested that Pop III.2 stars are
less massive ($M \lsim 40 - 60M_{\odot}$), being in the mass range
of ordinary iron core-collapse stars. Such stars explode and eject
heavy elements to contribute to chemical enrichment of the early
universe. The stars in this mass range are favorable candidates
for the elemental origin of extremely metal-poor stars in the
Galactic halo. We can explain why the signature of PISNe are not
seen with the senario that Pop III.1 stars are very massive, i.e.,
$M \gsim 300M_{\odot}$ and Pop III.2 stars are less massive, i.e.,
$M \lsim 40 - 60M_{\odot}$, although there is some uncertainty
with radiative feedback.

The possibility of our scenario that the majority of Pop III.1
stars is CVMSs ($M \gsim 300M_{\odot}$) and the majority of Pop
III.2 stars is ordinary massive stars ($M \lsim$ 40$M_{\odot}$) is
attractive because it explains the chemical evolution of the early
universe where the PISN mass range ($140M_{\odot} \leq M \leq
300M_{\odot}$) can be avoided. Moreover, the range we found for
large mass stars offers the attractive possibility that these
CVMSs are indeed progenitors of IMBHs. Although the presence of
IMBHs has not been firmly established, there are various recent
observational indications. If many of IMBHs indeed existed in the
early universe, even if they are rare today, that will provide
valuable insight to the formation of SMBHs and ultimately galaxy
formation and evolution.

\bigskip
TO, KN, and HU would like to thank R. Hix and F.-K. Thielemann for 
providing us with the nuclear reaction network. 
ST thanks M.J. Rees for valuable suggestions and discussion. This
work has been supported in part by World Premier International
Research Center Initiative, MEXT, and by the Grant-in-Aid for
Scientific Research of the JSPS (18104003, 18540231, 20540226,
20674003) and MEXT (17684008, 19047004, 20040004), Japan.

\begin{deluxetable}{ccc}
\tablecaption{Stellar evolution models with mass accretion rates as a func
stellar mass.
The mass accretion rates $dM_{\rm Y}/dt$ and $dM_{\rm YII}/dt$ are illustrated in
Figure~\ref{fig:dMYdt}. The mass accretion rates $dM_{\rm Mi}/dt$ are displayed
in Figure~\ref{fig:dMMdt}.
}
\tablewidth{0pt}
\tablehead{
\colhead{Models} &
\colhead{Mass accretion rate ($M_{\odot}{\rm yr^{-1}}$)} &
\colhead{}
}
\startdata
 {Y-1} &  $dM_{\rm Y}/dt$ & Equation~\ref{eqn:MdotY} \\
 {Y-2} &  $0.5 \times dM_{\rm Y}/dt$ & \\
 {Y-3} &  $0.33 \times dM_{\rm Y}/dt$ &\\
 {Y-4} &  $0.1 \times dM_{\rm Y}/dt$  &\\
 {Y-5} &  $0.05 \times dM_{\rm Y}/dt$ & \\
\hline
 {Y{\rm II}}  &  $dM_{\rm YII}/dt$ & Equation~\ref{eqn:MdotYII}\\
\hline
 {M-1} &  $dM_{\rm M1}/dt$ & Equation~\ref{eqn:MdotM1}\\
 {M-2} &  $dM_{\rm M2}/dt$ & Equation~\ref{eqn:MdotM2} \\
 {M-3} &  $dM_{\rm M3}/dt$ & Equation~\ref{eqn:MdotM3} \\
\enddata
\label{tab:ModelsY}
\end{deluxetable}

\begin{deluxetable}{ccccc}
\tablecaption{Stellar lifetime, final mass, CO core size, and final fate for each model.}
\tablewidth{0pt}
\tablehead{
\colhead{Models} &
\colhead{lifetime (yr)} &
\colhead{final mass $M_{\rm f}$ ($M_{\odot}$)} &
\colhead{CO core mass ($M_{\odot}$)} &
\colhead{final fate}
}
\startdata
 {Y-1} &   $2.3 \times 10^6$  &  915 & 370  & Core-Collpase \\
 {Y-2} &   $2.4 \times 10^6$  &  710 & 330  & Core-Collpase \\
 {Y-3} &   $2.5 \times 10^6$  &  610 & 275  & Core-Collpase \\
 {Y-4} &   $2.9 \times 10^6$  &  385 & 170  & Core-Collpase \\
 {Y-5} &   $3.1 \times 10^6$  &  275 & 123  & PISN \\
\hline
 {Y{\rm II}}  &   $5.5 \times 10^6$  &  40 & 14  & Core-Collpase \\
\hline
 {M-1} &   $2.2 \times 10^6$  &  321 & 155  & Core-Collpase  \\
 {M-2} &   $3.1 \times 10^6$  &  135 & 58 & Core-Collpase\\
 {M-3} &   $4.5 \times 10^6$  &  57 & 22  & Core-Collpase\\
\enddata
\label{tab:fates}
\end{deluxetable}



\onecolumn

\end{document}